\documentclass{article}
\usepackage{graphicx} % Required for inserting images
\usepackage{amsmath, amsfonts, amssymb}
\usepackage{authblk}

\title{\textbf{Security Implications and Mitigation Strategies in MPLS Networks}}
\author[1]{Ayush Thakur}
\affil[1]{\small Amity Institute of Information Technology, Amity University Noida, \texttt{ayush.th2002@gmail.com}}

\date{}

\begin{document}

\maketitle

\begin{abstract}
    Multiprotocol Label Switching (MPLS) is a high-performance telecommunications technology that directs data from one network node to another based on short path labels rather than long network addresses. Its efficiency and scalability have made it a popular choice for large-scale and enterprise networks. However, as MPLS networks grow and evolve, they encounter various security challenges. This paper explores the security implications associated with MPLS networks, including risks such as label spoofing, traffic interception, and denial of service attacks. Additionally, it evaluates advanced mitigation strategies to address these vulnerabilities, leveraging mathematical models and security protocols to enhance MPLS network resilience. By integrating theoretical analysis with practical solutions, this paper aims to provide a comprehensive understanding of MPLS security and propose effective methods for safeguarding network infrastructure.
\end{abstract}

\section{Introduction}

\subsection{Background}

Multiprotocol Label Switching (MPLS) represents a significant advancement in network engineering by enhancing packet forwarding efficiency and enabling sophisticated traffic management capabilities \cite{awduche2002internet}. MPLS operates by attaching labels to data packets, allowing routers to forward packets based on label values rather than complex network addresses. This approach facilitates faster data transfer, reduces latency, and supports Quality of Service (QoS) requirements, making it ideal for both service providers and enterprise networks \cite{armitage2000mpls}.

Despite these advantages, MPLS networks are not immune to security threats. The inherent characteristics of MPLS, such as label-based routing and label distribution, introduce unique vulnerabilities. Understanding these vulnerabilities and developing effective mitigation strategies are crucial for maintaining network security and performance.

\subsection{MPLS Network Architecture}

MPLS networks consist of several key components:
\begin{itemize}
    \item \textbf{Label Switch Routers (LSRs):} These routers are responsible for forwarding packets based on label values. LSRs perform label swapping, pushing, and popping operations as packets traverse the network \cite{armitage2000mpls}.
    \item \textbf{Label Edge Routers (LERs):} Positioned at the network edge, LERs handle packet ingress and egress. They assign labels to incoming packets and remove labels from outgoing packets \cite{yang2002edge}.
    \item \textbf{Label Distribution Protocols:} Protocols such as Label Distribution Protocol (LDP) \cite{guernsey2010security} and Resource Reservation Protocol with Traffic Engineering extensions (RSVP-TE) \cite{berger2003rfc3473} are used to distribute and manage labels within the network.
\end{itemize}

The forwarding process in MPLS networks can be represented using graph theory. Consider a network graph \( G = (V, E) \), where \( V \) represents nodes (LSRs and LERs) and \( E \) represents directed edges (network links). Each edge \( e \in E \) is associated with a label \( \ell \). The packet forwarding process can be modeled as a series of label swaps, where a packet follows a path \( P = (e_1, e_2, \ldots, e_n) \) through the network, with each edge \( e_i \) corresponding to a specific label.

\subsection{Security Challenges in MPLS Networks}

Several security challenges are inherent to MPLS networks, including:
\begin{itemize}
    \item \textbf{Label Spoofing:} Attackers may attempt to inject unauthorized labels into the network, disrupting normal packet forwarding \cite{raman2012mitigating}. Label spoofing can be modeled using probabilistic methods to analyze the likelihood of successful attacks based on label distribution protocols.
    \item \textbf{Traffic Interception:} MPLS does not inherently provide encryption or confidentiality. Attackers who gain access to MPLS paths can intercept and analyze traffic. This risk can be quantified using information theory to measure potential data exposure \cite{israr2015mpls}.
    \item \textbf{Denial of Service (DoS) Attacks:} Attackers can target MPLS network resources with excessive traffic, leading to degraded performance or network outages. DoS attacks can be modeled using queueing theory to assess their impact on network performance \cite{raju2013integrated}.
    \item \textbf{Misconfiguration and Vulnerabilities:} Misconfigurations in MPLS setups, such as improper label assignment or routing errors, can lead to security breaches \cite{rao2015traffic}. Mathematical modeling of network configurations can help identify potential vulnerabilities and misconfigurations.
\end{itemize}

The scope of this paper includes a theoretical analysis of MPLS security risks, an exploration of mathematical models for vulnerability assessment, and the development of practical mitigation strategies. The subsequent sections will delve deeper into these aspects, providing a comprehensive framework for addressing MPLS network security challenges.

\section{Security Implications in MPLS Networks}

\subsection{Label Spoofing}

Label spoofing in MPLS (Multiprotocol Label Switching) networks involves the unauthorized injection of labels, which can disrupt routing and misdirect traffic \cite{raman2012mitigating}. This threat can be particularly damaging, as it may lead to traffic being rerouted improperly or intercepted by unauthorized parties.

To understand label spoofing, we can model it as a perturbation in the label distribution protocol. Let \( \mathcal{L} \) denote the set of valid labels, and let \( \mathcal{L}_{att} \subset \mathcal{L} \) represent the set of spoofed labels. The probability \( P_{spoof} \) of a successful spoofing attack can be defined as:

\[
P_{spoof} = \Pr\left(\text{Spoofed Label Accepted} \mid \mathcal{L}_{att} \right)
\]

Assuming that labels are uniformly distributed, \( \mathcal{L} \) can be expressed as \( \mathcal{L} = \{ \ell_1, \ell_2, \ldots, \ell_m \} \), where \( m \) is the total number of labels. Under this assumption, the probability of an attacker successfully injecting a spoofed label is given by:

\[
P_{spoof} = \frac{|\mathcal{L}_{att}|}{m}
\]

where \( |\mathcal{L}_{att}| \) denotes the number of spoofed labels. To incorporate the dynamic nature of label distribution and switching, we extend the model to include label switching probabilities \( P(\ell_i \rightarrow \ell_j) \). The updated success rate of an attack can be modeled as:

\[
P_{spoof} = \sum_{i=1}^{m} \frac{|\mathcal{L}_{att} \cap \mathcal{L}_{i}|}{|\mathcal{L}_{i}|} \cdot \Pr(\text{Attack on Label } \ell_i)
\]

where \( \mathcal{L}_{i} \) represents the set of labels currently in use, and \( \Pr(\text{Attack on Label } \ell_i) \) is the probability of an attack on label \( \ell_i \).

Mitigation strategies can be employed to reduce the risk of label spoofing, which can be represented mathematically as follows:

\begin{itemize}
    \item \textbf{Authentication:} Implementing cryptographic authentication for label distribution. Let \( \mathcal{A}_L \) be the authentication function for labels, where each label \( \ell_i \) is associated with a cryptographic signature \( \sigma_i \). The security of the authentication process can be modeled by:

    \[
    P_{auth} = \Pr\left( \mathcal{A}_L(\ell_i, \sigma_i) \text{ is valid} \right)
    \]

    \item \textbf{Label Filtering:} Employing label filtering mechanisms to prevent unauthorized labels from being accepted. Let \( \mathcal{F}_L \) denote the filtering function. The effectiveness of label filtering can be expressed as:

    \[
    P_{filter} = \Pr\left( \text{Label } \ell_i \text{ is filtered out} \right)
    \]
\end{itemize}

\subsection{Traffic Interception}

MPLS networks are inherently vulnerable to traffic interception due to their lack of built-in encryption mechanisms. This vulnerability can expose sensitive data to unauthorized parties.

To quantify the risk of traffic interception, we can use Shannon's entropy \cite{zong2019analyzing}. Let \( H(X) \) represent the entropy of the information \( X \) being transmitted over MPLS paths. Entropy measures the amount of uncertainty or information contained in the data. The amount of intercepted information \( I \) is given by:

\[
I = H(X) \cdot \Pr(\text{Interception})
\]

where \( \Pr(\text{Interception}) \) is the probability of interception occurring. If \( S \) denotes the size of the secure data, the ratio \( R_{intercept} \) of intercepted information to the total secure data can be expressed as:

\[
R_{intercept} = \frac{I}{S} = \frac{H(X) \cdot \Pr(\text{Interception})}{S}
\]

A higher \( R_{intercept} \) value indicates a greater risk of information exposure, highlighting the potential severity of interception.

To mitigate the risk of traffic interception, the following strategies can be employed:

\begin{itemize}
    \item \textbf{Encryption:} Encrypting the data transmitted over the network can significantly reduce the risk of interception. Let \( \mathcal{E}(X) \) denote the encryption function applied to the data \( X \). The security of encryption is often quantified by the computational complexity required to break the encryption scheme. The probability that the encryption is secure can be modeled as:

    \[
    P_{encrypt} = \Pr\left( \mathcal{E}(X) \text{ remains secure} \right)
    \]

    where a higher \( P_{encrypt} \) signifies stronger encryption and reduced risk of interception.

    \item \textbf{Traffic Masking:} Employing traffic masking techniques can obscure the nature of the transmitted data and reduce the risk of traffic analysis. Let \( \mathcal{M}(X) \) represent the masking function applied to the data. The effectiveness of traffic masking can be represented by:

    \[
    P_{mask} = \Pr\left( \mathcal{M}(X) \text{ is untraceable} \right)
    \]

    Higher values of \( P_{mask} \) indicate more effective masking and a lower likelihood of successful traffic analysis.
\end{itemize}

\subsection{Denial of Service (DoS) Attacks}

DoS attacks target the availability of MPLS network resources, potentially leading to performance degradation or network outages \cite{genge2013analysis}. These attacks flood the network with excessive traffic, overwhelming its capacity and affecting its ability to process legitimate requests.

The impact of DoS attacks can be modeled using queueing theory \cite{wang2007queueing}. Let \( \lambda \) represent the arrival rate of attack traffic and \( \mu \) denote the service rate of the network resources. The traffic intensity \( \rho \) is given by:

\[
\rho = \frac{\lambda}{\mu}
\]

In this context, if \( \rho \) exceeds 1, the network becomes overloaded, and performance issues arise. The probability of packet loss \( P_{loss} \) in an \( M/M/1 \) queue, which is a basic model with a single server and Poisson arrival and service processes, can be approximated by:

\[
P_{loss} = 1 - \frac{1}{\rho}
\]

For more complex systems with multiple servers, the Erlang B formula is used to determine the blocking probability, which represents the probability that a new request will be lost due to all servers being busy. The Erlang B formula is:

\[
P_{loss} = \frac{\frac{(\lambda / \mu)^C}{C!}}{\sum_{k=0}^{C} \frac{(\lambda / \mu)^k}{k!}}
\]

where \( C \) is the number of servers. This formula provides a more accurate estimate of packet loss in multi-server systems.

Mitigation strategies for DoS attacks include:

\begin{itemize}
    \item \textbf{Rate Limiting:} Implementing rate limiting controls the rate of incoming traffic to prevent the network from being overwhelmed. The effectiveness of rate limiting can be expressed as:

    \[
    P_{rate\_limit} = \Pr\left( \mathcal{R}(\lambda) \text{ is effective} \right)
    \]

    where \( \mathcal{R}(\lambda) \) is the rate limiting function designed to maintain the traffic rate below a threshold that the network can handle.

    \item \textbf{Traffic Shaping:} Traffic shaping adjusts the flow of traffic to smooth out peaks and avoid sudden surges. The effectiveness of traffic shaping is represented by:

    \[
    P_{shape} = \Pr\left( \mathcal{T}(\lambda) \text{ prevents overload} \right)
    \]

    where \( \mathcal{T}(\lambda) \) is the traffic shaping function aimed at distributing traffic more evenly across the network to prevent spikes that could lead to overload.
\end{itemize}

\subsection{Misconfiguration and Vulnerabilities}

Misconfigurations in MPLS networks can create vulnerabilities that may be exploited, leading to various security risks and operational issues. Proper configuration management is crucial to maintain the security and reliability of the network.

The risk of vulnerabilities due to misconfigurations can be modeled using network reliability theory \cite{kondakci2015analysis}. Let \( R \) denote the reliability of the network configuration, and let \( M \) be the number of potential misconfigurations. The reliability \( R \) can be expressed as:

\[
R = \exp\left(-\frac{M}{N}\right)
\]

where \( N \) represents the total number of configuration parameters. This formula indicates that the network reliability \( R \) decreases as the number of misconfigurations \( M \) increases, highlighting the impact of errors or oversights in network configuration.

To mitigate the risks associated with misconfigurations, several strategies can be employed:

\begin{itemize}
    \item \textbf{Configuration Management:} Implementing robust configuration management practices is essential to prevent misconfigurations. The effectiveness of these practices can be represented by:

    \[
    P_{config\_manage} = \Pr\left( \mathcal{C}(M) \text{ effectively prevents misconfigurations} \right)
    \]

    where \( \mathcal{C}(M) \) denotes the configuration management function designed to minimize the risk of misconfigurations and ensure proper network setup.

    \item \textbf{Redundancy and Failover:} Designing network redundancy and failover mechanisms enhances reliability by providing backup components or systems that can take over in case of a failure. The reliability \( R_{red} \) of a redundant system can be calculated as:

    \[
    R_{red} = 1 - \prod_{i=1}^{k} (1 - R_i)
    \]

    where \( R_i \) is the reliability of each redundant component and \( k \) is the number of such components. This formula shows that the overall reliability \( R_{red} \) increases with the addition of more reliable redundant components, reducing the risk of downtime due to individual component failures.

\end{itemize}

\section{Mitigation Strategies for MPLS Network Security}

\subsection{Enhancing Label Security}

To address the vulnerabilities associated with label spoofing and ensure the integrity of label-based routing, several advanced mitigation strategies can be employed.

\subsubsection{Secure Label Distribution}

The security of label distribution protocols in MPLS networks can be significantly enhanced through the use of cryptographic techniques. Let \( \mathcal{L}_{auth} \) denote the set of authenticated labels, where each label \( \ell_i \) is associated with a cryptographic signature \( \sigma_i \). The process of authenticating these labels can be modeled using a secure key exchange protocol \( K \):

\[
\mathcal{L}_{auth} = \{ (\ell_i, \sigma_i) \mid \mathcal{V}(K, \ell_i, \sigma_i) = \text{true} \}
\]

Here, \( \mathcal{V} \) represents the verification function that checks whether the label \( \ell_i \) and its signature \( \sigma_i \) are valid under the key exchange protocol \( K \). The probability \( P_{auth} \) that a label is correctly authenticated is given by:

\[
P_{auth} = \Pr\left( \mathcal{V}(K, \ell_i, \sigma_i) = \text{true} \right)
\]

Incorporating digital signatures or message authentication codes (MACs) into the label distribution process ensures that each label is verified before being accepted by network devices. This approach mitigates the risk of label spoofing by ensuring that only authenticated labels are processed.

\begin{figure}[ht]
    \centering
    \includegraphics[width=\linewidth]{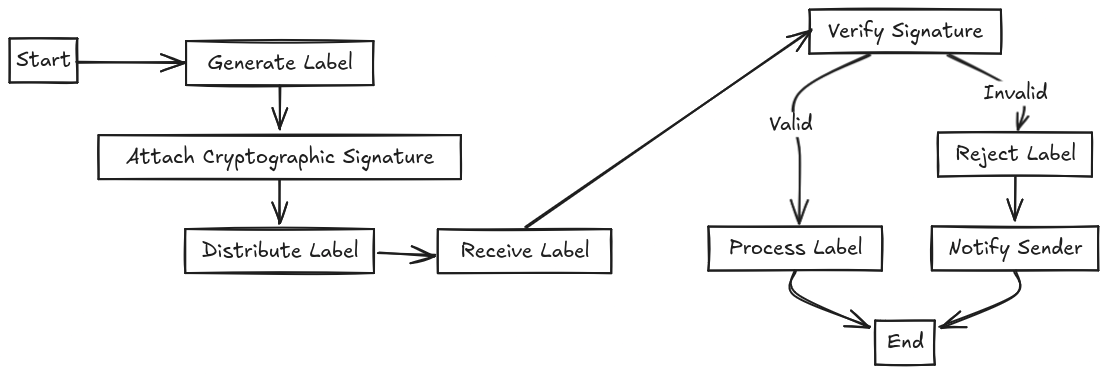}
    \caption{Label Distribution and Authentication Flowchart}
    \label{fig:Label Distribution and Authentication Flowchart}
\end{figure}

Figure \ref{fig:Label Distribution and Authentication Flowchart} illustrates the flow of the label distribution and authentication process. It shows how labels are generated, associated with cryptographic signatures, distributed, and verified, ensuring that only authorized labels are accepted by network devices.

% Mermaid Syntax
% flowchart TD
%     A[Start] --> B[Generate Label]
%     B --> C[Attach Cryptographic Signature]
%     C --> D[Distribute Label]
%     D --> E[Receive Label]
%     E --> F[Verify Signature]
%     F -- Valid --> G[Process Label]
%     F -- Invalid --> H[Reject Label]
%     H --> I[Notify Sender]
%     G --> J[End]
%     I --> J

\subsubsection{Label-Based Access Control}

Implementing label-based access control is crucial for securing MPLS networks by restricting the ability to inject or modify labels. This can be achieved through a label access matrix that defines which devices are allowed to access or handle specific labels.

Define a label access matrix \( \mathcal{A} \), where \( \mathcal{A}_{ij} \) indicates whether device \( i \) has access to label \( j \):

\[
\mathcal{A}_{ij} = \begin{cases} 
1 & \text{if device } i \text{ has access to label } j \\
0 & \text{otherwise}
\end{cases}
\]

In this matrix:
- \( \mathcal{A}_{ij} = 1 \) implies that device \( i \) is authorized to interact with label \( j \), including operations such as injection or modification.
- \( \mathcal{A}_{ij} = 0 \) indicates that device \( i \) is not authorized to interact with label \( j \).

To enforce this access control policy, network administrators can configure the network devices to check this matrix before allowing any interaction with labels. This ensures that only authorized devices can perform operations on specific labels, thereby preventing unauthorized label injection or modification and enhancing overall network security.

\subsection{Encryption and Confidentiality}

Securing data transmitted over MPLS networks requires robust encryption methods to protect against traffic interception.

\subsubsection{End-to-End Encryption}

End-to-end encryption ensures that data is encrypted at its source and decrypted only at its destination, providing a high level of security by preventing unauthorized access during transit. Let \( \mathcal{E}(X) \) represent the encryption function applied to the data \( X \), and \( \mathcal{D}(X) \) denote the decryption function. The security of the encrypted data can be expressed as:

\[
P_{secure} = \Pr\left( \mathcal{D}(\mathcal{E}(X)) = X \right)
\]

This probability indicates how reliably the original data \( X \) can be recovered after decryption, assuming the encryption and decryption functions are correctly implemented.

To further enhance security, encryption algorithms such as the Advanced Encryption Standard (AES) are widely used. AES is known for its robustness and efficiency. The security strength of AES is often evaluated based on the length of the encryption key \( k \), with the security strength \( \mathcal{S}(k) \) defined as:

\[
\mathcal{S}(k) = 2^{k}
\]

where \( k \) is the key length in bits. Larger key sizes offer greater resistance to brute-force attacks, thereby providing stronger security. For instance, a 256-bit key provides significantly more security than a 128-bit key.

\begin{figure}[ht]
    \centering
    \includegraphics[width=0.7\linewidth]{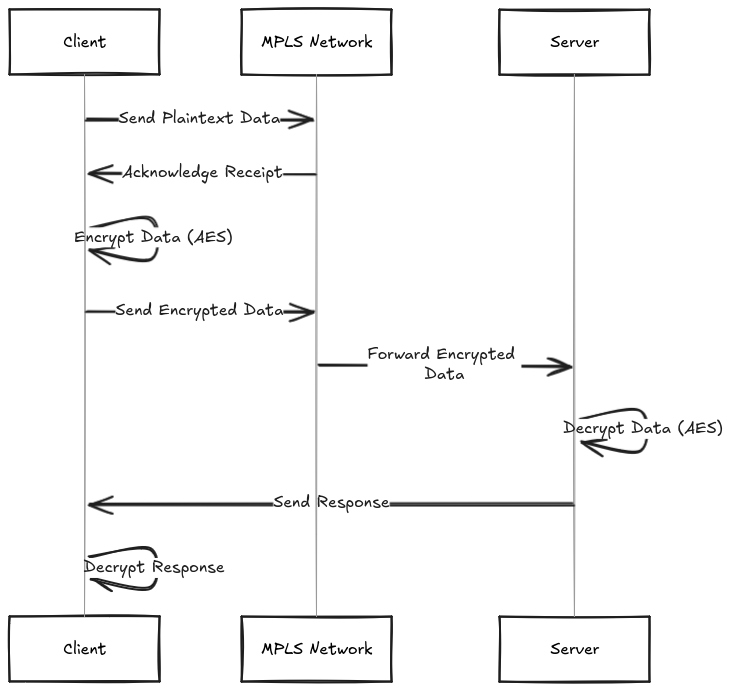}
    \caption{Traffic Encryption Protocol Sequence Diagram}
    \label{fig:Traffic Encryption Protocol Sequence Diagram}
\end{figure}

Figure \ref{fig:Traffic Encryption Protocol Sequence Diagram} illustrates the sequence of operations involved in end-to-end encryption, showing how data is encrypted at the source and decrypted at the destination to ensure secure communication.

% Mermaid
% sequenceDiagram
%     participant Client
%     participant MPLS Network
%     participant Server

%     Client->>MPLS Network: Send Plaintext Data
%     MPLS Network->>Client: Acknowledge Receipt
%     Client->>Client: Encrypt Data (AES)
%     Client->>MPLS Network: Send Encrypted Data
%     MPLS Network->>Server: Forward Encrypted Data
%     Server->>Server: Decrypt Data (AES)
%     Server->>Client: Send Response
%     Client->>Client: Decrypt Response

\subsubsection{Traffic Encryption Protocols}

Deploying encryption protocols such as IPsec within MPLS networks enhances security by adding a layer of protection for data in transit \cite{simatimbe2020performance}. To quantify the security provided by IPsec, consider the following:

Let \( \mathcal{T}(X) \) denote the traffic encryption function, and \( \mathcal{I}(X) \) represent the integrity check function. When applied together, these functions ensure that the traffic is both encrypted and verified for integrity. The combined security provided by IPsec can be expressed as:

\[
P_{ipsec} = \Pr\left( \mathcal{T}(\mathcal{I}(X)) \text{ is intact} \right)
\]

Here:
- \( \mathcal{I}(X) \) performs an integrity check on the data \( X \), ensuring that it has not been altered.
- \( \mathcal{T}(\mathcal{I}(X)) \) represents the result of encrypting the integrity-checked data, thus protecting it from unauthorized access and ensuring its confidentiality.

The probability \( P_{ipsec} \) measures the likelihood that the encrypted and integrity-checked traffic remains intact and unaltered, reflecting the effectiveness of IPsec in safeguarding data.

\subsection{Mitigating DoS Attacks}

To combat DoS attacks, implementing strategies that control and manage network traffic is crucial.

\subsubsection{Rate Limiting and Traffic Policing}

Rate limiting is an effective strategy for managing traffic rates and preventing network overloads. Let \( \mathcal{R}(\lambda) \) denote the rate limiting function, where \( \lambda \) represents the actual traffic rate and \( \lambda_{max} \) is the maximum allowable rate. The rate limiting function is defined as:

\[
\mathcal{R}(\lambda) = \min(\lambda, \lambda_{max})
\]

This function ensures that the traffic rate does not exceed \( \lambda_{max} \), effectively controlling the traffic flow and preventing overload.

The effectiveness of rate limiting in managing traffic can be assessed by comparing actual traffic rates to the allowable limits. The probability \( P_{limit} \) that rate limits are enforced effectively is given by:

\[
P_{limit} = \Pr\left( \lambda \leq \lambda_{max} \right)
\]

This probability measures how often the actual traffic rate remains within the allowed limit, indicating the effectiveness of the rate limiting mechanism in preventing traffic overload.

\begin{figure}[ht]
    \centering
    \includegraphics[width=0.5\linewidth]{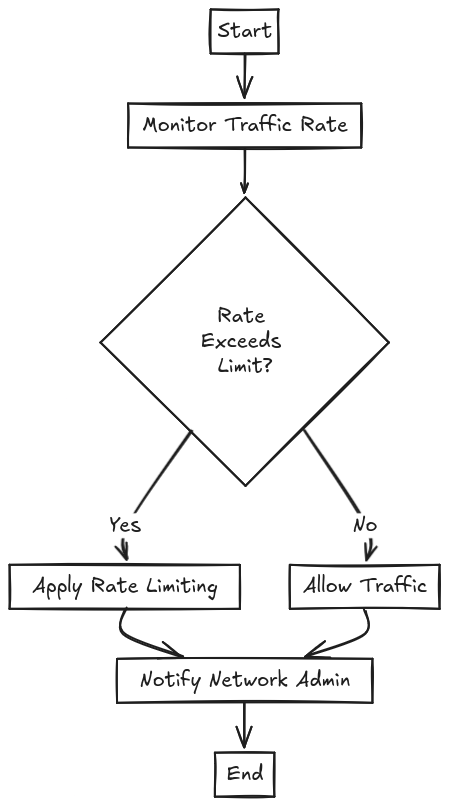}
    \caption{DoS Attack Mitigation Flowchart}
    \label{fig:DoS Attack Mitigation Flowchart}
\end{figure}

Figure \ref{fig:DoS Attack Mitigation Flowchart} provides a visual representation of the rate limiting and traffic policing process. It illustrates the steps involved in monitoring traffic rates, applying rate limits when necessary, and notifying network administrators, thereby helping to mitigate Denial of Service (DoS) attacks and manage network traffic efficiently.

% Mermaid
% flowchart TD
%     A[Start] --> B[Monitor Traffic Rate]
%     B --> C{Rate Exceeds Limit?}
%     C -- Yes --> D[Apply Rate Limiting]
%     C -- No --> E[Allow Traffic]
%     D --> F[Notify Network Admin]
%     E --> F
%     F --> G[End]

\subsubsection{Traffic Shaping}

Traffic shaping is a technique used to smooth out traffic flows and prevent sudden spikes, thereby enhancing network stability and performance \cite{erramilli2002self}. By regulating the rate at which traffic is sent or received, traffic shaping helps in maintaining a consistent traffic pattern and can mitigate the effects of Denial of Service (DoS) attacks.

Define \( \mathcal{S}(T) \) as the traffic shaping function applied over a time interval \( T \). The effectiveness \( P_{shape} \) of traffic shaping in reducing traffic burstiness can be modeled as:

\[
P_{shape} = \Pr\left( \text{Traffic follows expected profile} \mid \mathcal{S}(T) \right)
\]

In this context:
- \( \mathcal{S}(T) \) represents the application of traffic shaping over the specified time interval \( T \), where traffic is adjusted to conform to a predetermined profile.
- \( P_{shape} \) denotes the probability that the traffic adheres to the expected profile after shaping, indicating the effectiveness of the traffic shaping strategy in preventing sudden traffic bursts.

\subsection{Configuration Management and Network Reliability}

To address misconfigurations and vulnerabilities, robust configuration management and reliability measures are essential.

\subsubsection{Automated Configuration Management}

Automated configuration management tools are essential for detecting and correcting configuration issues in network systems. Let \( \mathcal{M}(C) \) denote the configuration management function, where \( C \) represents the current configuration state. The effectiveness \( P_{config} \) of this function in maintaining a correct configuration can be expressed as:

\[
P_{config} = \Pr\left( \mathcal{M}(C) \text{ maintains correct configuration} \right)
\]

In this context:
- \( \mathcal{M}(C) \) is responsible for managing and verifying the configuration \( C \) to ensure it aligns with desired security policies and operational standards.
- \( P_{config} \) indicates the probability that the configuration management process successfully maintains an accurate and secure configuration.

Regular audits and automated verification processes are integral to configuration management, ensuring adherence to security policies and minimizing the risk of vulnerabilities arising from misconfigurations.

\subsubsection{Network Redundancy and Failover}

Implementing network redundancy involves designing failover mechanisms to enhance system reliability and maintain service availability. Let \( R_{sys} \) represent the reliability of a redundant system with \( k \) components, where each component \( i \) has a reliability \( R_i \). The overall reliability \( R_{red} \) of the redundant system can be calculated as:

\[
R_{red} = 1 - \prod_{i=1}^{k} (1 - R_i)
\]

In this formula:
- \( R_i \) is the reliability of each individual component in the system.
- The expression \( \prod_{i=1}^{k} (1 - R_i) \) represents the probability that all components fail.
- Therefore, \( R_{red} \) gives the probability that at least one component is functioning, thus ensuring system reliability.

Redundancy helps ensure that the failure of one or more components does not lead to a total system outage, thereby maintaining overall network reliability and availability.

\section{Implementation and Practical Considerations}

\subsection{Implementation of Security Measures}

Implementing the discussed security measures in MPLS networks involves several practical steps:

\subsubsection{Deploying Secure Label Distribution}

To deploy secure label distribution, network operators should integrate cryptographic authentication mechanisms into their label distribution protocols. This includes configuring Label Distribution Protocol (LDP) \cite{aboul2001qos} or Resource Reservation Protocol (RSVP) \cite{zhang1993rsvp} with extensions for digital signatures and message authentication codes (MACs) \cite{katz2008aggregate}.

\subsubsection{Implementing End-to-End Encryption}

For end-to-end encryption, the network should incorporate encryption protocols such as IPsec to secure data transmitted over MPLS paths. This involves:

\begin{itemize}
    \item \textbf{Selecting Encryption Protocols:} Choosing appropriate encryption standards (e.g., AES) based on the required security level and performance trade-offs.
    \item \textbf{Configuring IPsec:} Implementing IPsec policies for data encryption and integrity across MPLS connections.
\end{itemize}

\subsubsection{Applying Rate Limiting and Traffic Shaping}

To manage DoS attacks, network operators should configure rate limiting and traffic shaping mechanisms:

\begin{itemize}
    \item \textbf{Rate Limiting:} Setting thresholds for traffic rates on network devices to prevent excessive traffic from overwhelming the network.
    \item \textbf{Traffic Shaping:} Implementing traffic profiles and smoothing techniques to manage traffic bursts and maintain consistent flow.
\end{itemize}

\subsubsection{Enhancing Configuration Management}

Effective configuration management requires:

\begin{itemize}
    \item \textbf{Automated Tools:} Utilizing automated configuration management tools to detect and rectify misconfigurations.
    \item \textbf{Regular Audits:} Conducting periodic audits to ensure configurations adhere to security policies and best practices.
\end{itemize}

\subsection{Practical Considerations}

\subsubsection{Resource Allocation}

Implementing these security measures may require additional resources, including computational power and network bandwidth. Operators must balance security enhancements with the network’s performance requirements. Resource allocation strategies should include:

\begin{itemize}
    \item \textbf{Scalability:} Ensuring that security measures scale with network growth.
    \item \textbf{Performance Impact:} Evaluating the impact of security protocols on network performance and optimizing configurations to minimize overhead.
\end{itemize}

\subsubsection{Training and Awareness}

Training network personnel on security best practices and the operation of advanced security features is crucial. Operators should:

\begin{itemize}
    \item \textbf{Conduct Training Programs:} Provide training on cryptographic protocols, encryption techniques, and configuration management.
    \item \textbf{Raise Awareness:} Foster awareness of potential security threats and the importance of maintaining secure configurations.
\end{itemize}

\subsubsection{Compliance and Standards}

Adhering to industry standards and regulatory requirements is essential for ensuring comprehensive security. Compliance with standards such as ISO/IEC 27001 \cite{humphreys2016implementing} for information security management and NIST guidelines can help:

\begin{itemize}
    \item \textbf{Ensure Best Practices:} Align security measures with established best practices and standards.
    \item \textbf{Facilitate Audits:} Simplify compliance with regulatory audits and assessments.
\end{itemize}

\section{Conclusion}

In conclusion, securing MPLS networks involves a multifaceted approach that addresses label spoofing, traffic interception, DoS attacks, and misconfigurations through advanced mathematical models and practical implementations. By integrating secure label distribution, encryption, traffic management, and robust configuration practices, network operators can significantly enhance the security and reliability of MPLS networks. The successful implementation of these strategies requires careful planning, resource allocation, and adherence to industry standards to ensure both network performance and security are optimized. As MPLS networks continue to evolve, ongoing vigilance and adaptation of security measures will be essential to counter emerging threats and maintain robust network integrity.

\bibliographystyle{alpha}
\bibliography{ref}

\end{document}